\documentstyle[psfig]{l-aa}

\newcommand{\beqa}{\begin{eqnarray*}}
\newcommand{\eeqa}{\end{eqnarray*}}
\newcommand{\beqan}{\begin{eqnarray}}
\newcommand{\eeqan}[1]{\label{#1}\end{eqnarray}}
\newcommand{\beq}{\begin{equation}}
\newcommand{\eeq}{\end{equation}}

\newcommand{\lp}{ \left(}
\newcommand{\rp}{ \right)}

\begin{document}

\bibliographystyle{plain}

\thesaurus{
02.08.1; 
08.09.3, 
08.18.1. 
}

\title{The hot side of the lithium dip -- LiBeB abundances 
beyond the main sequence}

\author{Corinne Charbonnel \inst{1},  Suzanne Talon \inst{2,3}}

\offprints{Corinne Charbonnel ; corinne@obs-mip.fr} 

\institute{
Laboratoire d'Astrophysique de Toulouse, CNRS UMR5572, OMP,
14, Av. E.Belin, 31400 Toulouse, France 
\and 
D\'epartement de Physique, Universit\'e de Montr\'eal, Montr\'eal PQ H3C 3J7, Canada
\and
CERCA, 5160 boul. D\'ecarie, Montr\'eal PQ H3X 2H9, Canada
}

\date{Received 30 July 1999 / Accepted 13 September 1999}

\maketitle
\markboth{Charbonnel \& Talon
- The hot side of the Li dip : LiBeB beyond the main sequence}{}

\begin{abstract}
We extend to the case of A and early-F type stars our study of the 
transport of matter and angular momentum by wind-driven meridional 
circulation and shear turbulence. 
We show that our fully consistent treatment of the same hydrodynamical 
processes which can account for C and N anomalies in B type stars 
(Talon et al. 1997) and for the shape of the hot side of the Li dip 
in the open clusters (Talon \& Charbonnel 1998)
also explains LiBeB observations in stars with T$_{\rm eff}$ higher 
than 7000 K on the main sequence 
as well as in their evolved counterparts.

\keywords{Li dip; hydrodynamics; turbulence; 
Stars: interiors, rotation, abundances}

\end{abstract}

\section{Lithium abundance on the hot side of the dip}
Due to its fragility to nuclear reactions,
lithium is a very powerful tracer of the physical conditions in stellar
interiors. During the last two decades, numerous abundance determinations 
of Li for both field and cluster low mass stars have indeed brought to light 
the occurrence of transport processes of chemicals in 
stellar radiative zones.

One of the most striking signatures of this (these) mechanism(s) 
is the drop-off in the lithium content of main sequence F-stars in
a range of 300 K in effective temperature centered around 6700 K, first
discovered in the Hyades by Wallerstein et al. (1965) and later
confirmed by Boesgaard \& Tripicco (1986). 
This feature appears in all galactic clusters older than $\sim$ 200 Myrs 
as well as in field stars
(see Michaud \& Charbonneau 1991 and Balachandran 1995 for references). 

Relatively few abundance determinations are available for main sequence
cluster stars on the hot side of this so-called lithium dip. 
The scarceness of data is due to the fact that lithium can be
observed in slow rotators only ($V\sin i$ lower than $\sim$ 
70 km.sec$^{-1}$), while most of the stars on the hot side of the dip are 
actually fast rotators. 
In the Hyades, Praesepe and Coma\footnote{which all have approximately 700 Myrs},
these objects show lithium abundances close to the galactic value 
(log$N$(Li)$\simeq$3.31, with log$N$(H)=12), 
except for a few Li deficient 
Am-stars (Boesgaard 1987, Burkhart \& Coupry 1989, 1998\footnote{In
Praesepe, Am stars present the same Li abundance as F stars on the hot
side of the lithium dip.}, 1999).
In the field, the occurrence of Li-underabundant main sequence (or
slightly evolved) stars originating from the hot side of the dip 
is not rare (Balachandran 1990, Burkhart \& Coupry 1991).
For these objects, the lithium abundance shows indeed a dispersion of one 
order of magnitude below the galactic value.

Among the different processes which have been proposed to account for
the lithium distribution 
in Pop I low-mass stars\footnote{Gravitational 
settling and radiative diffusion (Michaud 1986, 
Proffitt \& Michaud 1991, Richer \& Michaud 1993), mass loss 
(Hobbs et al. 1989, Guzik et al. 1987, Schramm et al. 1990, Swenson \& 
Faulkner 1992), meridional circulation (Charbonneau \& Michaud 1988, 1990, 
1991), turbulent mixing induced by rotation (Vauclair et al. 1978, Vauclair
1988, Pinsonneault et al. 1989, 1990, Deliyannis \& Pinsonneault 1990,
Charbonneau \& Michaud 1991, Charbonnel et al. 1992, 1994, Charbonnel \&
Vauclair 1992, Chaboyer et al. 1995, Talon \& Charbonnel 1998), 
gravity waves (Garcia Lopez \& Spruit 1991, Montalban \& Schatzman 1994).},  
atomic diffusion is the only one which can be calculated from first 
principles without any arbitrary parameter. 
Its importance has been spectacularly demonstrated in the Sun, thanks to
helioseismology (Bahcall \& Pinsonneault 1995, Christensen-Dalsgaard et
al. 1996, Richard et al. 1996). 
The nature and strength of the abundance anomalies (for lithium, but
also for helium and metals) due to diffusion alone are expected to change 
in the domain going from G to F and A stars due to the shallowing of 
the superficial H-He convective zone when $T_{\rm eff}$ increases and to 
the competition between radiative acceleration and gravitational settling. 
It has however been known for a long time 
(Michaud 1970, Michaud et al. 1976, Vauclair et al. 1978) 
that diffusion theory actually predicts abundance anomalies much larger 
than the ones observed, revealing the occurrence of some perturbating
macroscopic process(es). New calculations performed by the Montreal
group taking into account turbulent diffusion yield results 
in better agreement with observations, once more illustrating the need for 
an improved treatment of hydrodynamical processes 
(Richer et al. 1999a).

Regarding the F stars in the Li-dip region, carbon, oxygen 
and boron under-abundances are indeed expected in the case of pure
diffusion (Michaud 1986, Turcotte et al. 1998), but are not observed in 
the Hyades members (Boesgaard 1989, Friel \& Boesgaard 1990, Garcia
Lopez et al. 1993) nor in the Be-deficient field stars (Boesgaard et al. 1977). 
In addition, in this framework Li settles and remains in a buffer zone
below the convective envelope and should be dredged-up after the
turnoff; this is not observed in M67 subgiant stars 
which originate from the dip
(Pilachowski et al. 1988, Balachandran 1995, Deliyannis et al. 1997). 
This implies that another process competes with diffusion and leads to
nuclear destruction of lithium inside the dip. 
Finally, a macroscopic process is expected to limit both the predicted 
Li overabundances due to radiative acceleration in stars with 
T$_{\rm eff}\simeq$7000 K (Michaud 1986), and the predicted strong
Li underabundances due to settling for stars with T$_{\rm eff}>$7200 K 
(Richer \& Michaud 1993), which are not observed in open clusters.
The overabundance problem is less crucial 
compared to previous studies with the new 
radiative force calculations by Richer et al. (1999b), who follow the
strong coupling that exists between the atomic diffusion of one element
and the abundance of another one (Richer et al. 1997).
Indeed, in these stars the radiative acceleration on lithium is never 
higher than gravity, but of the same order of magnitude.
However, the behavior of the more massive stars remains puzzling.

Important clues for understanding the macroscopic processes that compete 
with atomic diffusion in A and F stars come from lithium observations in 
subgiant stars originating from the hot side of the dip. 
Looking at subgiants rather than giants allows to track signatures of 
mechanisms acting deep in the star on the main sequence while permitting 
to avoid further complications due to possible lithium destruction in more 
advanced stages (see Charbonnel 1995). 
Observations by Alschuler (1975) of a few field stars crossing for the first 
time the Hertzsprung gap indicated that lithium depletion starts earlier than 
predicted by standard dilution for stars more massive than 2 M$_{\odot}$.
Later on, significant lithium depletion was observed in a non negligible
number of slightly evolved field stars (Brown et al. 1989, Balachandran 1990). 
Recently, Dias et al. (1999) studied the behavior of lithium in an homogeneous 
sample of field subgiants observed by L\`ebre et al. (1999) for which Hipparcos 
data allowed the precise determination of both the evolutionary status and 
the mass. They confirmed that stars originating from the hot side of the dip
present a large range of lithium abundances which can not be explained by 
standard dilution alone and which reflect different degrees of depletion of 
this element while on the main sequence, even if its signature does not 
appear at the stellar surface at the age of the Hyades (see Vauclair 1991, 
Charbonnel \& Vauclair 1992). 
In this cluster, dilution is not sufficient to explain lithium values in 
evolved stars (which have masses of the order of 2.2 M$_{\odot}$), while 
main sequence stars present galactic abundance (Boesgaard et al. 1977, 
Duncan et al. 1998). 
This has been observed also in open clusters with turnoff masses higher
than $\sim$ 1.6 M$_{\odot}$ (Gilroy 1989). 

In Talon \& Charbonnel (1998, hereafter TC98), we assessed 
the role of the wind-driven meridional circulation and of shear turbulence as
described by Zahn (1992) and Maeder (1995) in the transport of chemicals 
and angular momentum in low mass stars. We showed that the shape of the 
hot side of the Li dip in the Hyades was successfully explained within this 
framework which also reproduces the C and N anomalies in B-type stars 
(Talon et al. 1997). 

In the present paper we study the impact of rotational mixing in stars 
hotter than 7000 K, and extend our computations up to the completion of the 
first dredge-up in order to confront our predictions with observations in 
subgiant and giant stars.
In \S 2 we recall the input physics of our models as well as the
observational constraints on the evolution of the surface velocity.
The global characteristics of the models are discussed in \S 3.  
We then present the theoretical predictions for the evolution of the
LiBeB and helium abundances, and compare our results with observations for 
both field and cluster evolved stars in \S 4.

\section{Models including rotational mixing and atomic diffusion}
\subsection{The transport processes}
Our models (computed with the Toulouse-Geneva evolutionary code) include 
both element segregation and rotation-induced mixing,
and we treat simultaneously the transport of matter and angular
momentum. The internal rotation profile evolves self-consistently under
the action of meridional circulation, which is treated as a truly
advective process, as described by Zahn (1992)
and of shear stresses, treated as a turbulent viscosity
(cf. Talon \& Zahn 1997).
As mentioned in TC98, each of these two processes involves a free
parameter of order unity. Here, we did not calibrate the parameters
and simply used the values which had been used in previous calculations
(TC98) namely $C_h=1$ and $\nu_{\rm th}=2/5$.
We recall that the first free parameter describes
the weakening effect of the horizontal turbulence on the vertical
transport of chemical (cf. Zahn 1992 for details)
and the second free parameter describes the
weakening of the restoring forces of the density stratification
on the development of the vertical shear. The vertical turbulent
viscosity $\nu_v$ is then
\beq
\nu _v = \nu_{\rm th} K \lp \frac{r}{N} \frac{\partial
\Omega}{\partial
r} \rp ^2,
\label{dift}
\eeq
where
$N$ is the Brunt-V\"ais\"al\"a frequency and $K$ is 
the thermal diffusivity.
We refer to TC98 (see also Talon et al. 1997) for more details. 

We do treat the atomic diffusion of H, He, CNO, Ne and Mg. The
diffusion coefficients are computed with Paquette et al. (1986)
prescription, and we include the diffusion due to gravity and to the 
concentration and thermal gradients.
The case of the LiBeB elements is peculiar, since on the hot side of the 
dip they are (partly) supported by radiative acceleration which is of the 
same order of magnitude 
as gravity (cf. Richer et al. 1999b). 
We thus decided to compute their evolution due to the macroscopic
mixing only. 

\subsection{Constraints on the evolution of the surface rotation}
The observational data indicate that, at the age of
the Hyades, stars on the hot side of the dip have kept their initial
rotation velocities (due to their very shallow convection envelope), while 
cooler stars have already been spun down by a magnetic torque.  
To take into account the large dispersion of the observed $V\sin i$ in the
mass range we consider, we computed models with different initial velocities:
100 km.sec$^{-1}$, which corresponds to the mean velocity of the Hyades
stars with $T_{\rm eff}>$7000 K, and 50 and 150 km.sec$^{-1}$ (see
Gaig\'e 1993 and TC98). 
When the star leaves the main sequence, the evolution of its surface
rotation velocity in our models is due only to the stellar expansion
and momentum redistribution,
i.e., we do not treat magnetic braking. This has no influence on the
light element evolution on the subgiant branch since
during this phase the surface depletion
is largely dominated by dilution. 

\subsection{Input physics and stellar masses}
The stellar models presented here include the same physics as in
TC98. The radiative opacities from Iglesias \& Rogers (1996) are
complemented by the low-temperature opacities Alexander \& Ferguson (1995). 
The thermonuclear reaction rates are from Caughlan \& Fowler (1988). 

We chose $z = 0.20$ as the initial metallicity of our
models in order to compare our results
to observations in Pop I stars in general\footnote{In TC98 the initial 
mixture was suited for precise comparison to the Hyades cluster.}. 
The initial helium content is 0.28. The relative
ratios for the heavy elements correspond to the mixture by Grevesse \&
Noels (1993), and the isotopic ratios are from Maeder (1983).

We performed our computations for three stellar masses, namely 1.5,
1.85 and 2.2 M$_{\odot}$. At the age of the Hyades, the first one has 
an effective temperature at the edge of the Li-dip, while the last one 
corresponds to a slightly evolved star. 

In these models, we assumed that no mass loss takes place. Indeed, the
evolution of the rotation velocities of A and 
F stars indicates that the matter leaving the stars provides no
(significant) torque to their surface. The small amount of mass lost at
the surface thus leaves the star with its own angular momentum.
Taking this into account in the numerical simulation 
would only marginally alter the result and we thus neglected it.

\section{Global characteristics of the models}
\subsection{HR diagram}

\begin{figure}
\psfig{figure=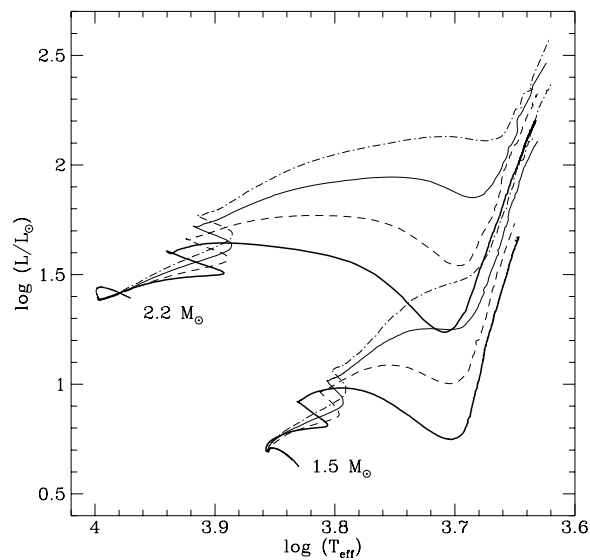,height=8cm}
\caption{Evolution of the models in the HRD from the zero age
main sequence up to the completion of the first dredge-up
(only two masses are shown for clarity). 
The solid,
dashed and dashed-dotted lines correspond respectively to the stars with
initial rotation velocities of 100, 50 and 150 km.sec$^{-1}$. The bold
tracks are followed by non-rotating models
\label{fig_dhrcomplet}}
\end{figure}

\begin{table}
\caption{Characteristics of the stellar models at the end of the main
sequence}
\begin{center}
\begin{tabular}{ccccc}\hline
\multicolumn{1}{c}{M$_*$/M$_{\odot}$} &
\multicolumn{1}{c}{V} &
\multicolumn{1}{c}{t} &
\multicolumn{1}{c}{log T$_{\rm eff}$} &
\multicolumn{1}{c}{log L/L$_{\odot}$} \\ 
\multicolumn{1}{c}{} &
\multicolumn{1}{c}{(km.sec$^{-1}$)} &
\multicolumn{1}{c}{(Gyrs)} &
\multicolumn{1}{c}{} &
\multicolumn{1}{c}{} \\ \hline
1.5 & 150 & 2.52 & 3.797 & 1.078 \\
    & 100 & 2.39 & 3.802 & 1.026 \\
    &  50 & 2.24 & 3.808 & 0.975 \\
    &   0 & 2.18 & 3.818 & 0.961 \\ \hline
1.85 & 150 & 1.30  & 3.846 & 1.46  \\
     & 100 & 1.25  & 3.851 & 1.412 \\
     &  50 & 1.17  & 3.860 & 1.355 \\
     &   0 & 1.13  & 3.870 & 1.323 \\ \hline
2.2 & 150 & 0.784 & 3.902 & 1.788 \\
    & 100 & 0.744 & 3.905 & 1.737 \\
    &  50 & 0.709 & 3.913 & 1.676 \\
    &   0 & 0.691 & 3.917 & 1.633 \\
\hline
\end{tabular}
\end{center}
\end{table}

\begin{figure}[h]
\centerline{
\psfig{figure=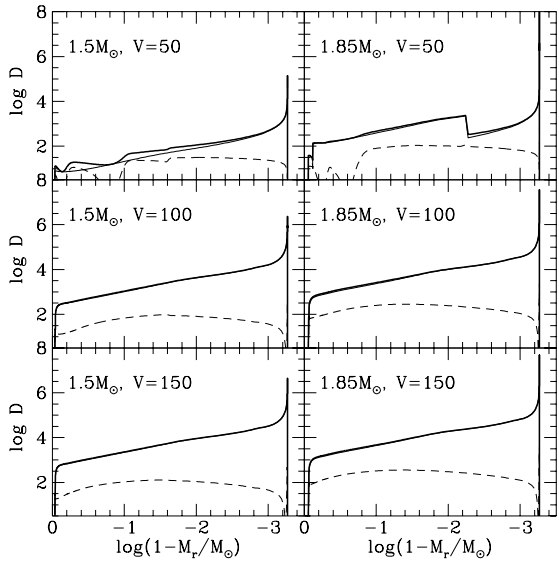,height=8cm}
}
\caption{ Diffusion coefficient profiles at the age of the 
Hyades for the 1.5 and 1.85 M$_{\odot}$ models 
and for the three different velocities. 
The bold line represents the total diffusion coefficient, the thin full
line shows the turbulent diffusion coefficient, and the dashed line, the
effective diffusion coefficient related to the meridional circulation. 
In all cases, turbulence dominates the transport of the chemicals. 
After the main sequence, the abundance evolution is largely dominated
by the dredge-up
\label{fig_profilscd_Hyades}}
\end{figure}

Fig.~\ref{fig_dhrcomplet} displays the evolutionary paths followed by our 
models in the HR diagram\footnote{Let us notice that, on the scale of
our HR diagram, the structural changes due to rotation are not visible
on the ZAMS.
Indeed, rotating models have lower luminosities and effective temperatures
due to the reduction of gravity close to the equator by the centrifugal force, 
leading also to longer main sequence lifetimes. 
Here however, for the 1.5 M$_{\odot}$ rotating at 150 km.sec$^{-1}$, the 
difference in effective temperature on the ZAMS in only 30 K while the 
lifetime is lengthened (due to structural effects only) by about 15 Myrs.}. 
In Table 1, we give their main characteristics at the end of the main sequence. 
Values for standard models (with no rotation) are also quoted. 

The differences of global characteristic between the models including
mixing and the standard models is striking.
Indeed, as a consequence of mixing, the abundance profiles inside the star are
modified compared to the standard ones. In particular, the slight
changes in the helium profile (see Fig.~3 and \S 3.3) affect the evolution.
Furthermore,
since mixing keeps adding fresh fuel to the core, the lifetime on the 
main sequence increases with the rotation rate.
At the turnoff, the faster models are cooler and more luminous than the 
models with slow or no rotation. 
When the star leaves the main sequence, the differences in opacity between 
the models lead to evolution at higher luminosity on the subgiant branch
for the faster rotators (see \S 3.3).

\subsection{Transport coefficients inside the stars}
\begin{figure}[ht]
\centerline{
\psfig{figure=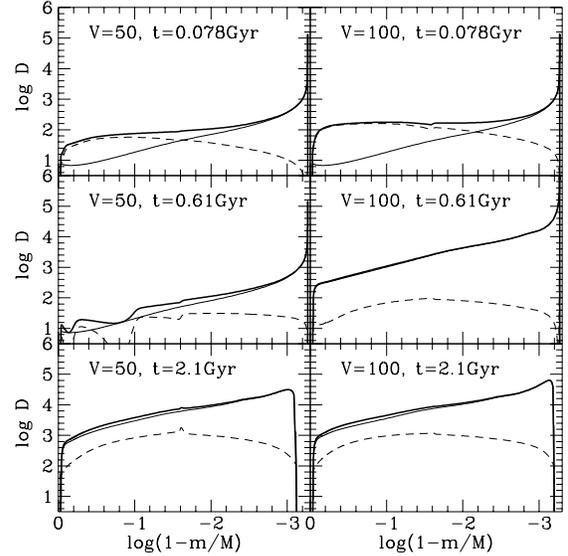,height=8cm}
}
\caption{Diffusion coefficient profiles at three ages the 1.5 M$_{\odot}$ 
model for two different velocities.
The line symbols are the same as in Fig~\ref{fig_profilscd_Hyades}
\label{fig_profilscd_1p5_V50V100}}
\end{figure}

Fig.~\ref{fig_profilscd_Hyades} shows the diffusion coefficients inside 
the models which are still on the main sequence at the age of the Hyades.
In Fig.~\ref{fig_profilscd_1p5_V50V100} we follow the evolution of these
coefficients for the 1.5 M$_{\odot}$ star with the lowest rotation 
velocities. 
During the first part of the evolution ($t=78$ Myrs and also 
$t=610$ Myrs for the slowest model), turbulence has not appeared yet
and the full thin line represents only the lower limit of viscosity,
namely radiative viscosity. Then settling of helium is possible, as
can be seen in Figs.~\ref{fig_evolutionab1p5},
\ref{fig_evolutionab1p85} and \ref{fig_evolutionab2p2}. During that first
period, meridional circulation advects momentum towards the interior,
building up the differential rotation. When it becomes sufficient
and for the rest of the main sequence evolution, turbulent diffusion 
dominates the transport of chemicals.

\begin{figure*}
\centerline{
\psfig{figure=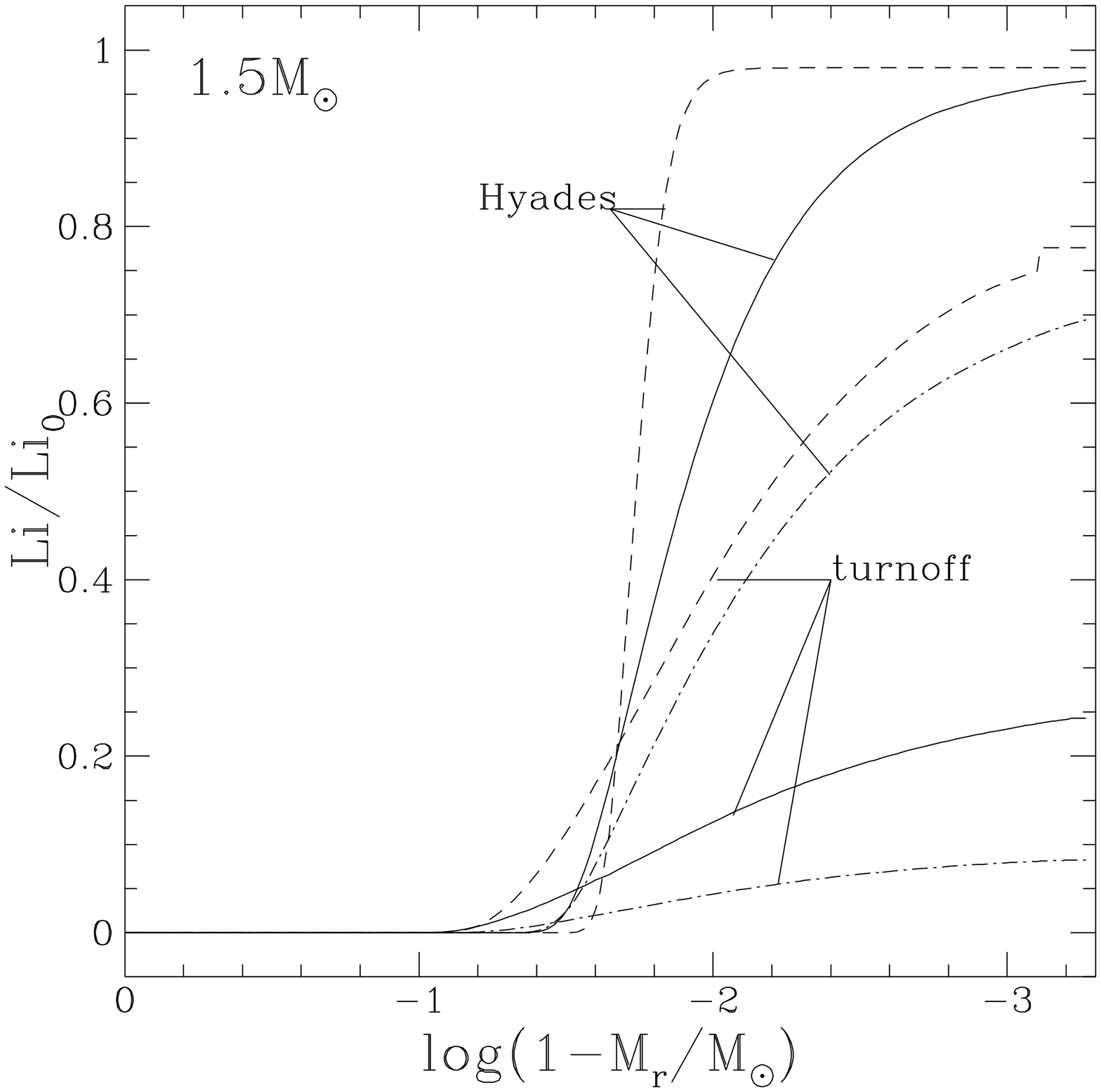,height=6cm}
\psfig{figure=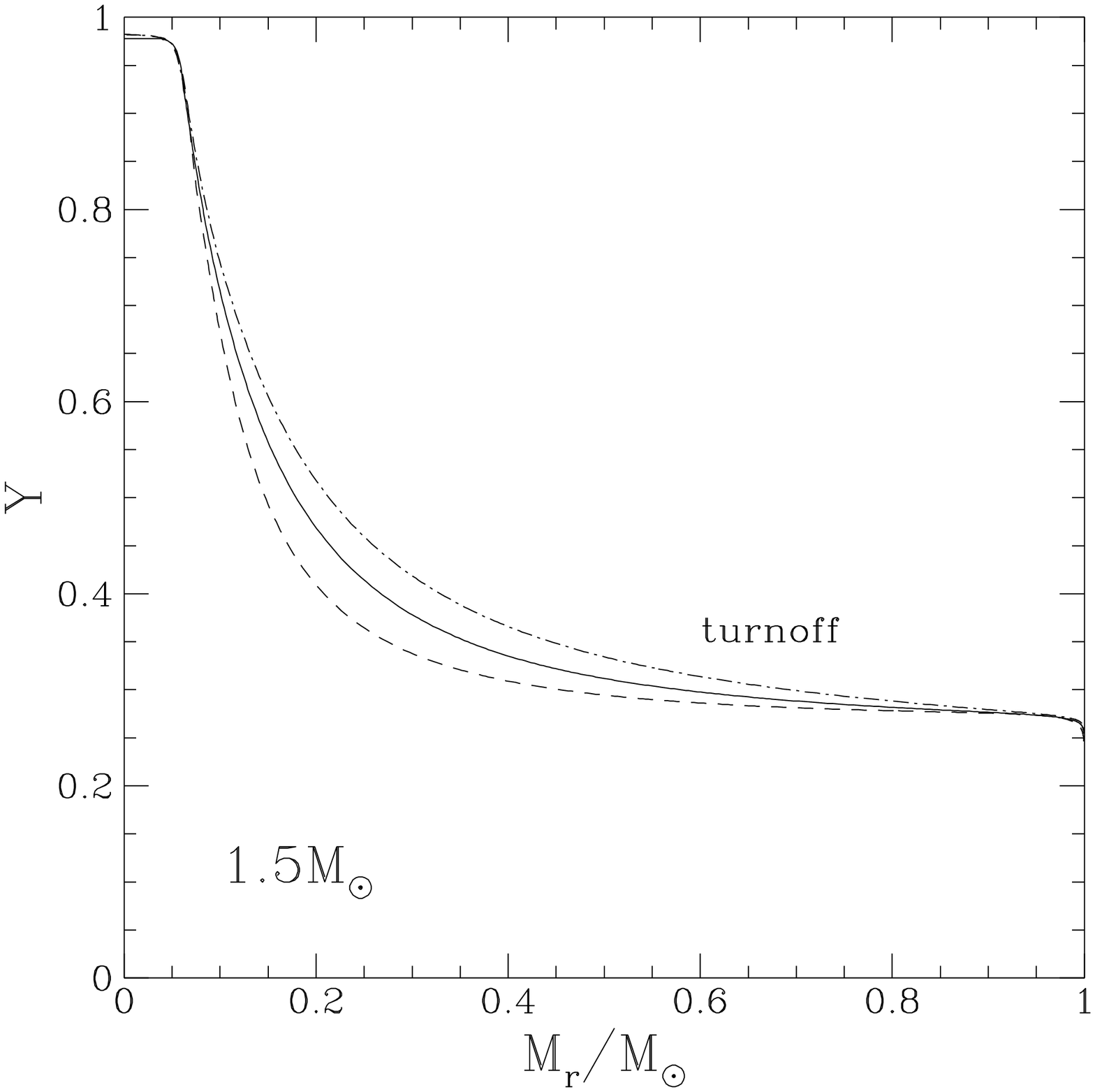,height=6cm}}
\centerline{
\psfig{figure=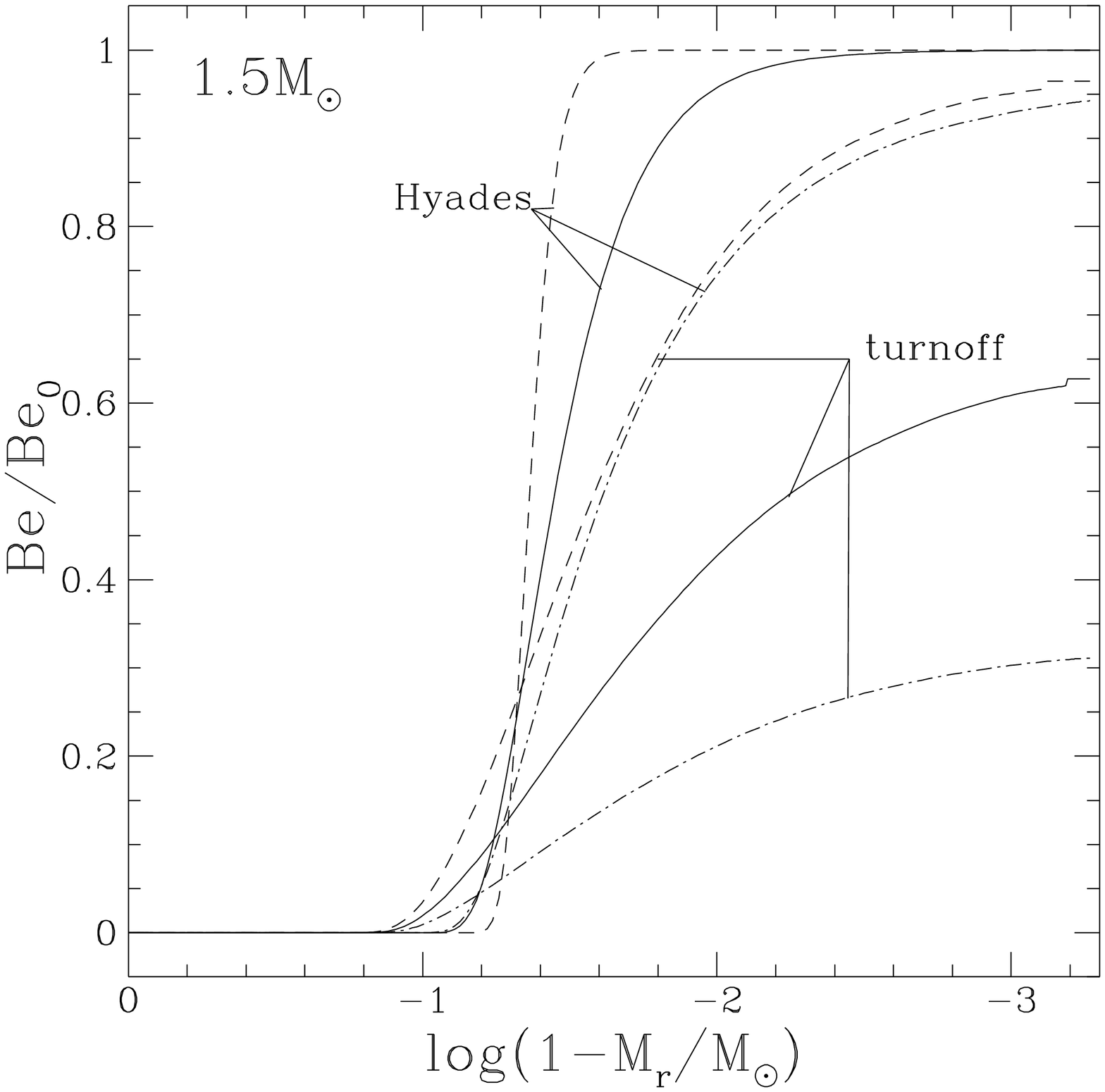,height=6cm}
\psfig{figure=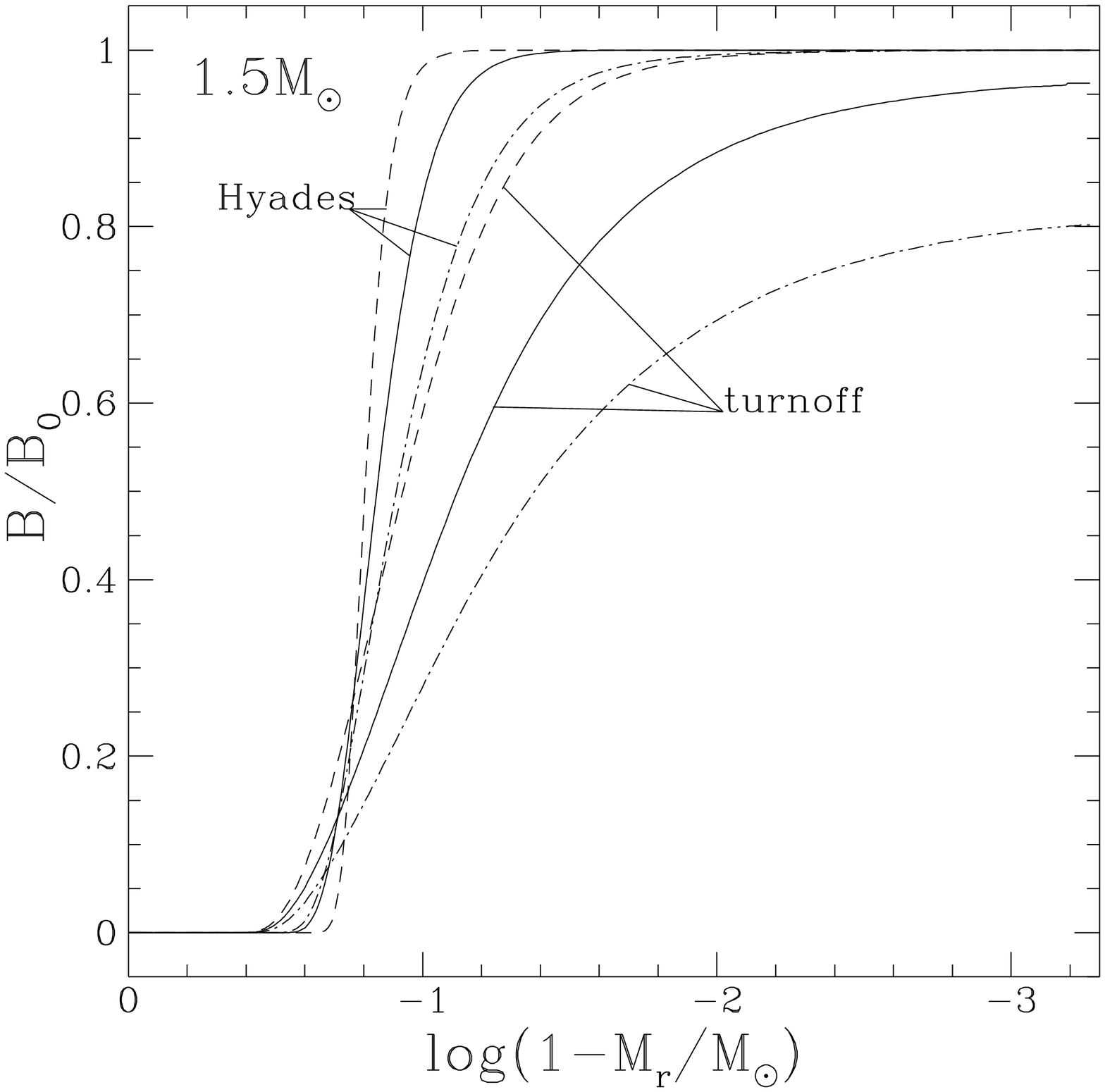,height=6cm}
}
\caption{
Abundance profiles of LiBeB inside the 1.5 M$_{\odot}$ 
star at the age of the Hyades and at the turnoff. The profiles of helium
(in mass fraction) are shown at the turnoff only. The solid, dashed and
dashed-dotted lines correspond respectively to the initial rotation
velocities of 100, 50 and 150 km.sec$^{-1}$ \label{fig_abprof1p5}}
\end{figure*}

The situation is different for angular momentum though.
Indeed, according to observations, main sequence 
stars hotter than the Li-dip are not slowed down by a magnetic torque. 
As we discussed in TC98, since the complete equation for the transport of 
angular momentum admits a stationary solution, these stars soon reach a 
regime with no net angular momentum flux. 
The rotation profile results from the equilibrium of the advection of 
angular momentum by meridional circulation and of the turbulent diffusion 
due to shear turbulence\footnote{Remember we make the hypothesis that the 
major source of turbulence is shear.} which counterbalance each other. 
This stresses the need for a fully consistent treatment of the 
hydrodynamical processes present in stars.

\subsection{Abundance profiles}
In Fig.~\ref{fig_abprof1p5} we show the interior abundance profiles of the 
LiBeB elements in the 1.5 M$_{\odot}$ model, at the age of the Hyades and at 
the turnoff for the three initial equatorial velocities. 
In the slowest model, the abundance profiles are relatively steep in the
region of destruction of the LiBeB elements by nuclear reactions. 
More rapid rotation leads to the destruction of these 
light elements in a larger region. 

\begin{figure*}
\centerline{
\psfig{figure=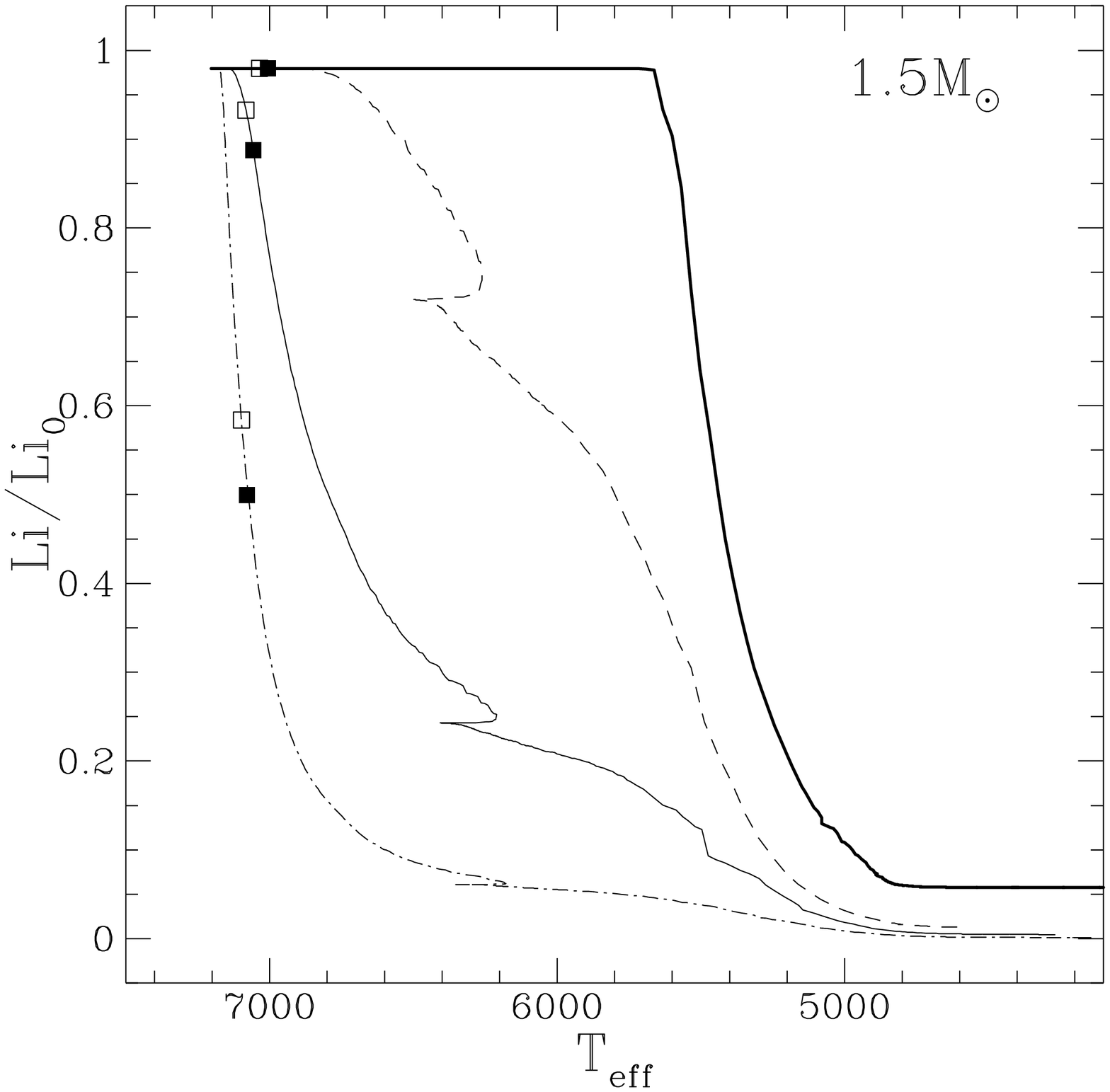,height=6cm}
\psfig{figure=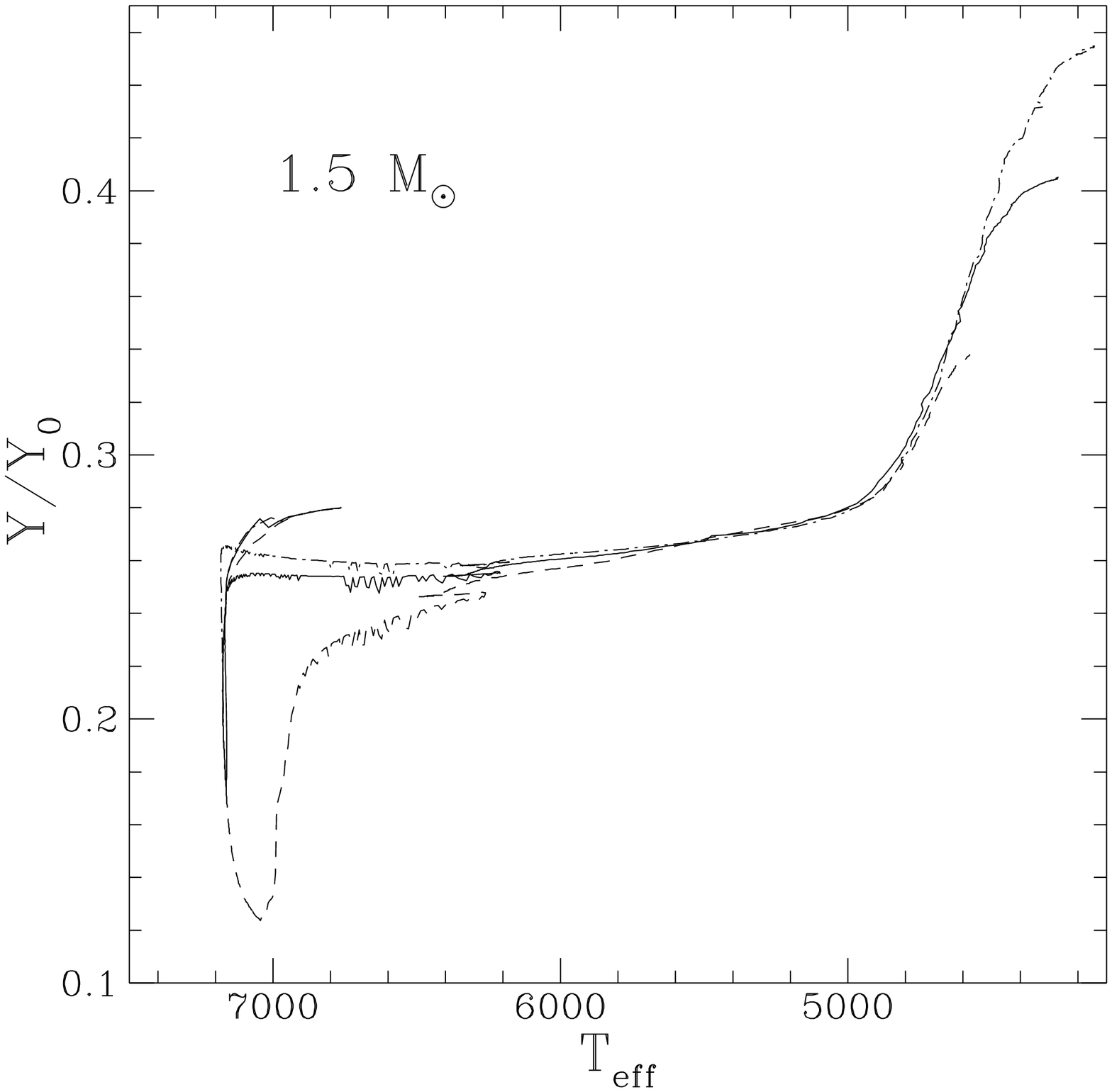,height=6cm}
}
\centerline{
\psfig{figure=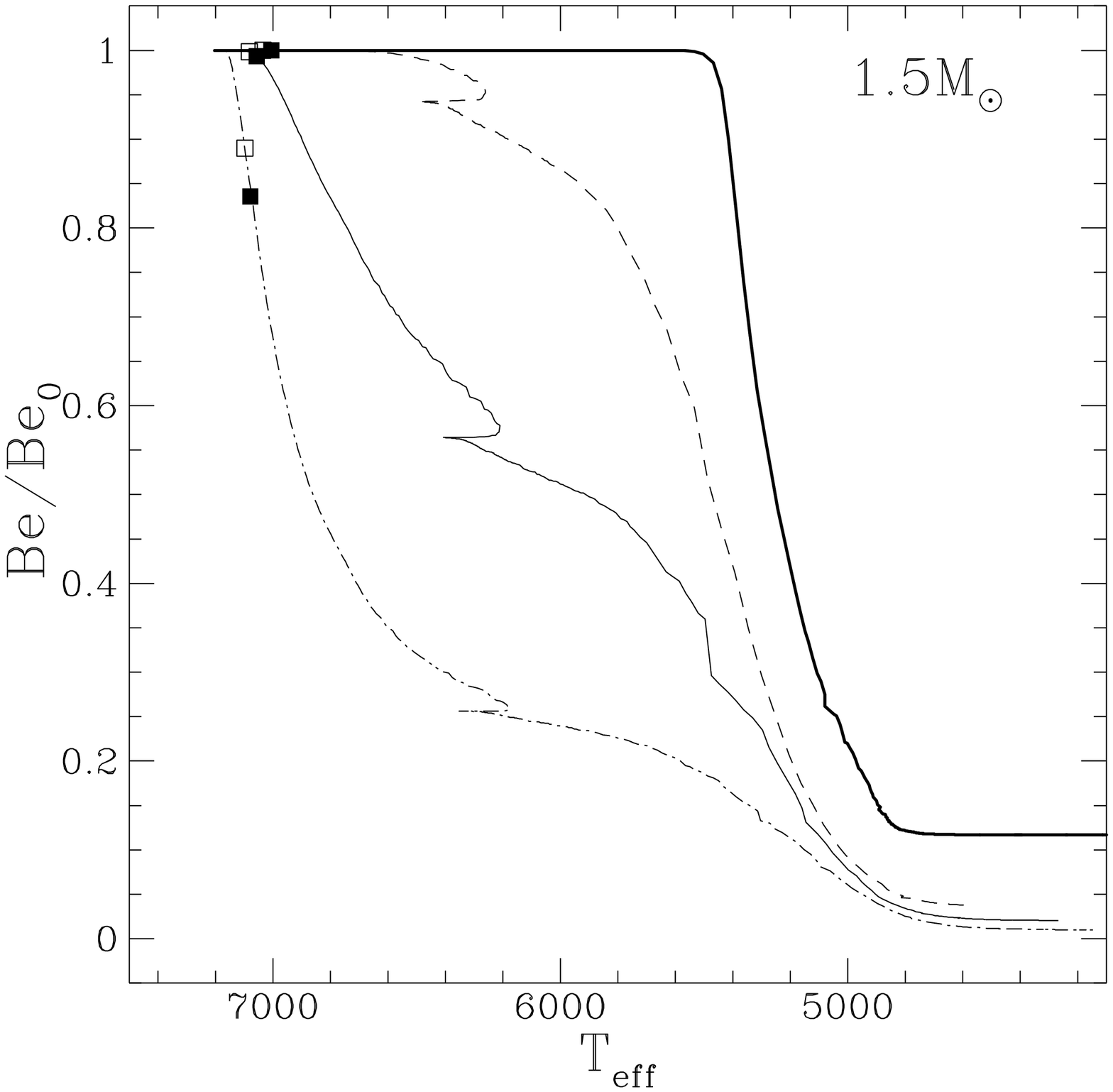,height=6cm}
\psfig{figure=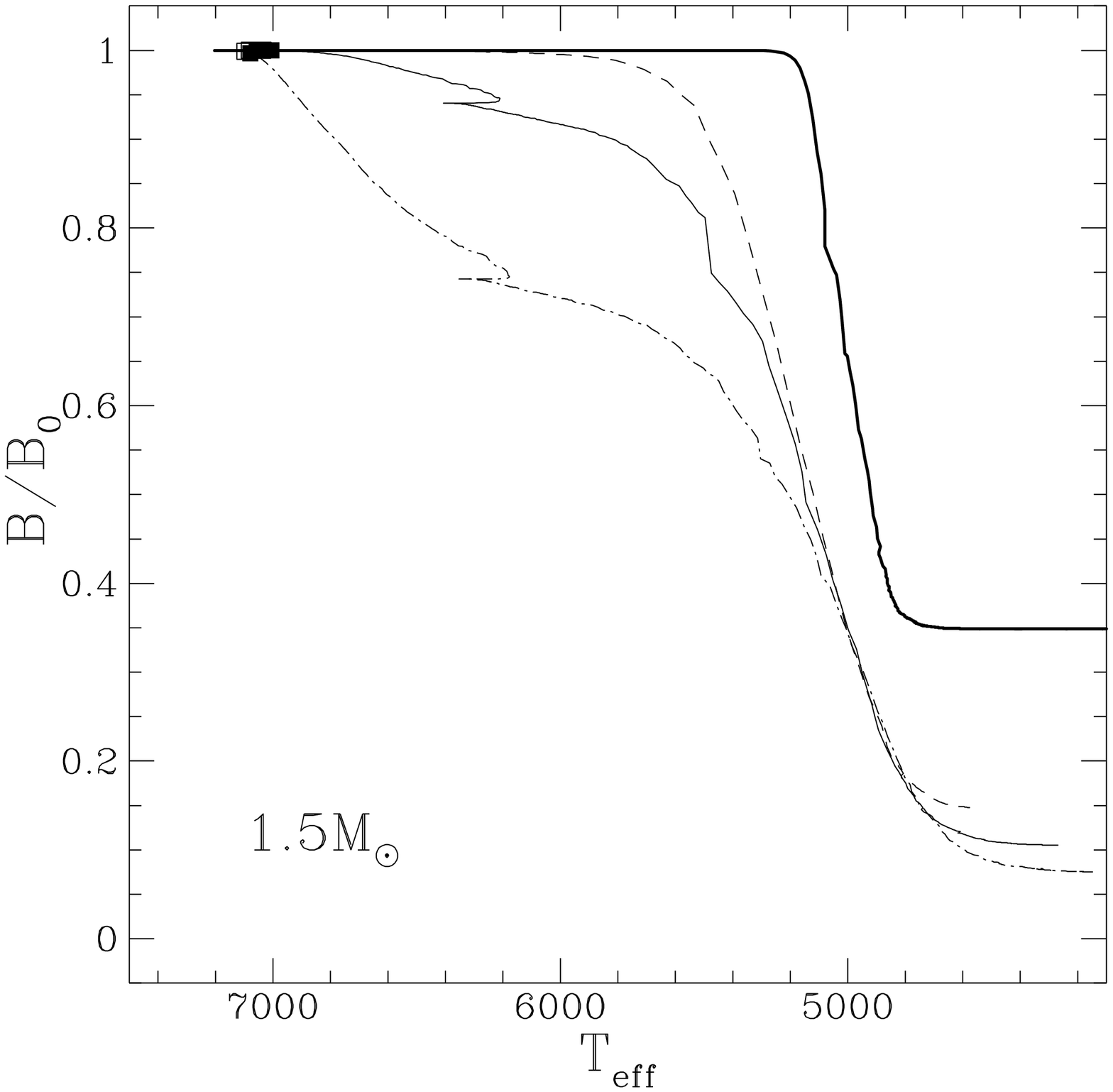,height=6cm}
}
\caption{Evolution of the LiBeB elements and of the helium mass
fraction at the surface of the 1.5 M$_{\odot}$ star as a function of its
effective temperature. The line symbols are as in Fig.~1. The open and black
squares indicate the evolutionary point at 0.7 and 0.8 Gyr respectively
\label{fig_evolutionab1p5}}
\end{figure*}

\begin{figure*}
\centerline{
\psfig{figure=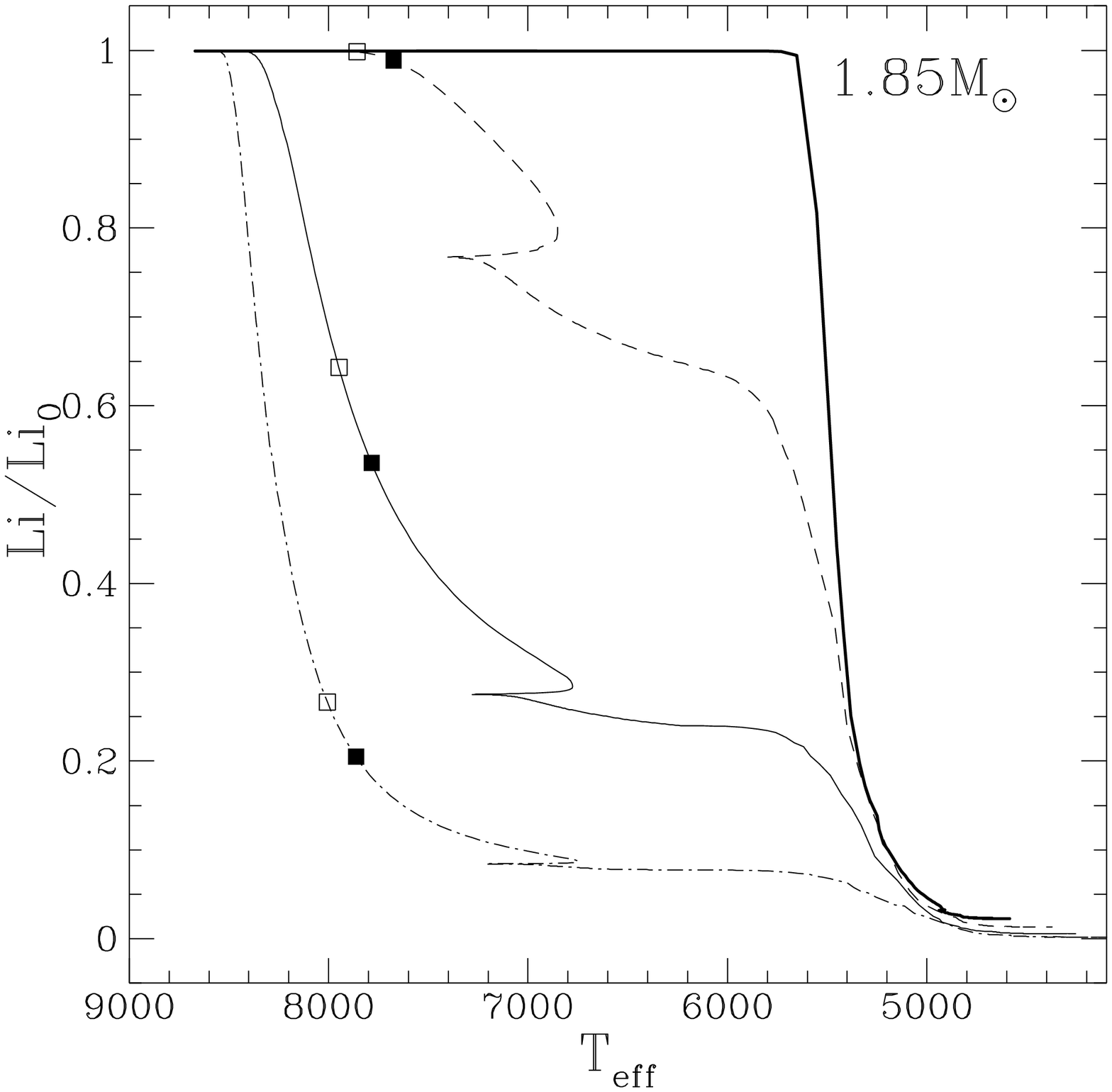,height=6cm}
\psfig{figure=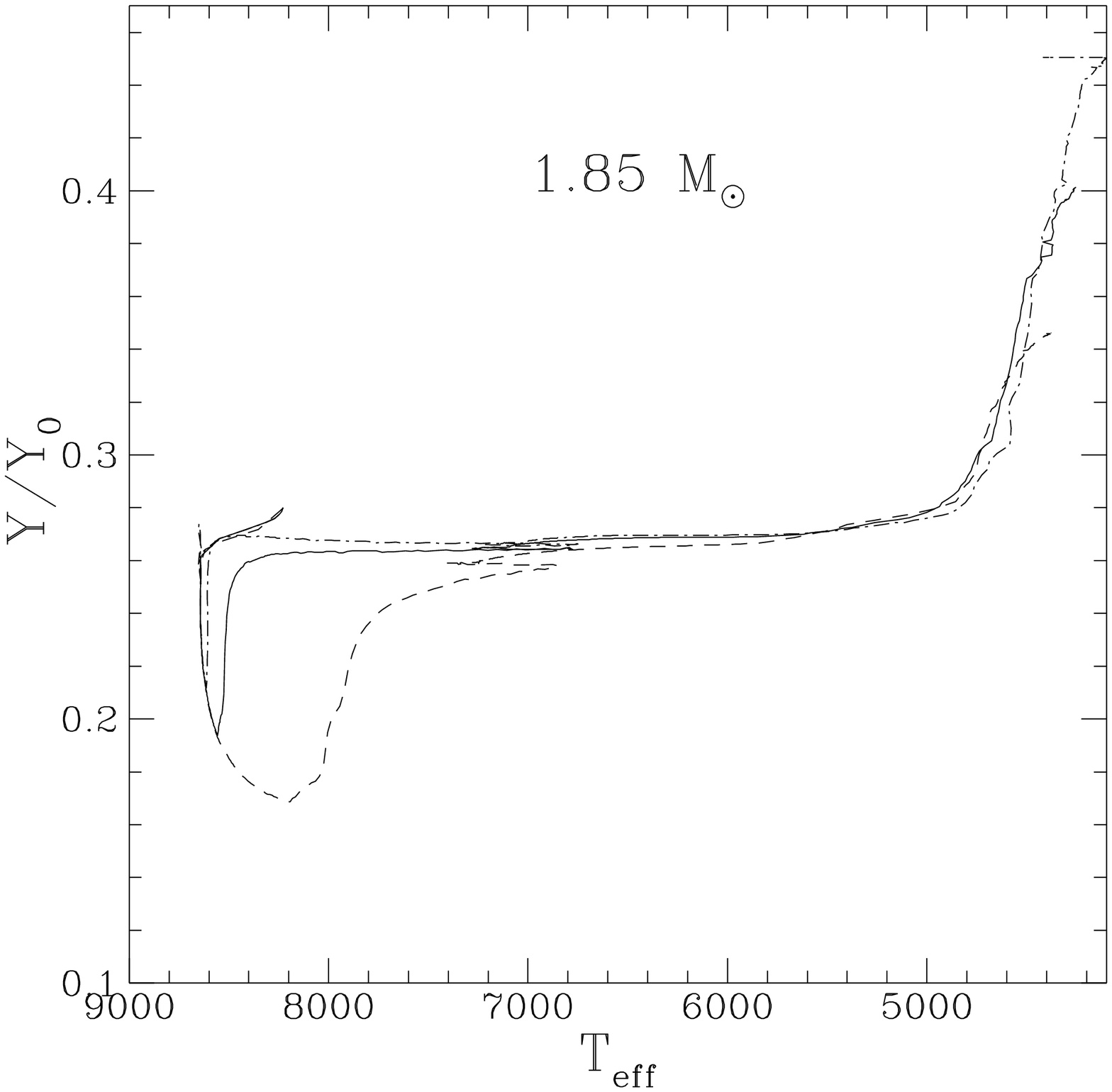,height=6cm}
}
\centerline{
\psfig{figure=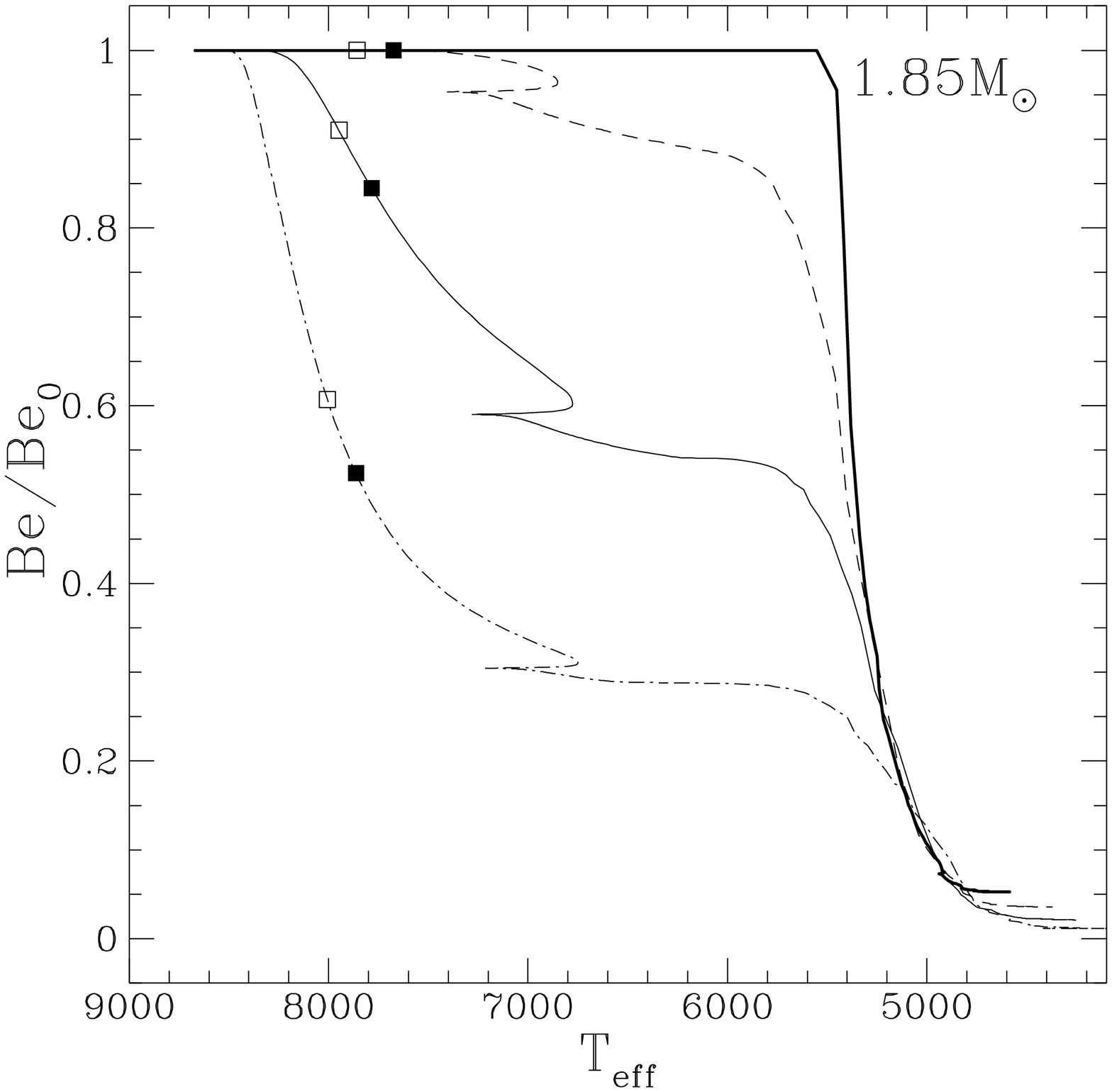,height=6cm}
\psfig{figure=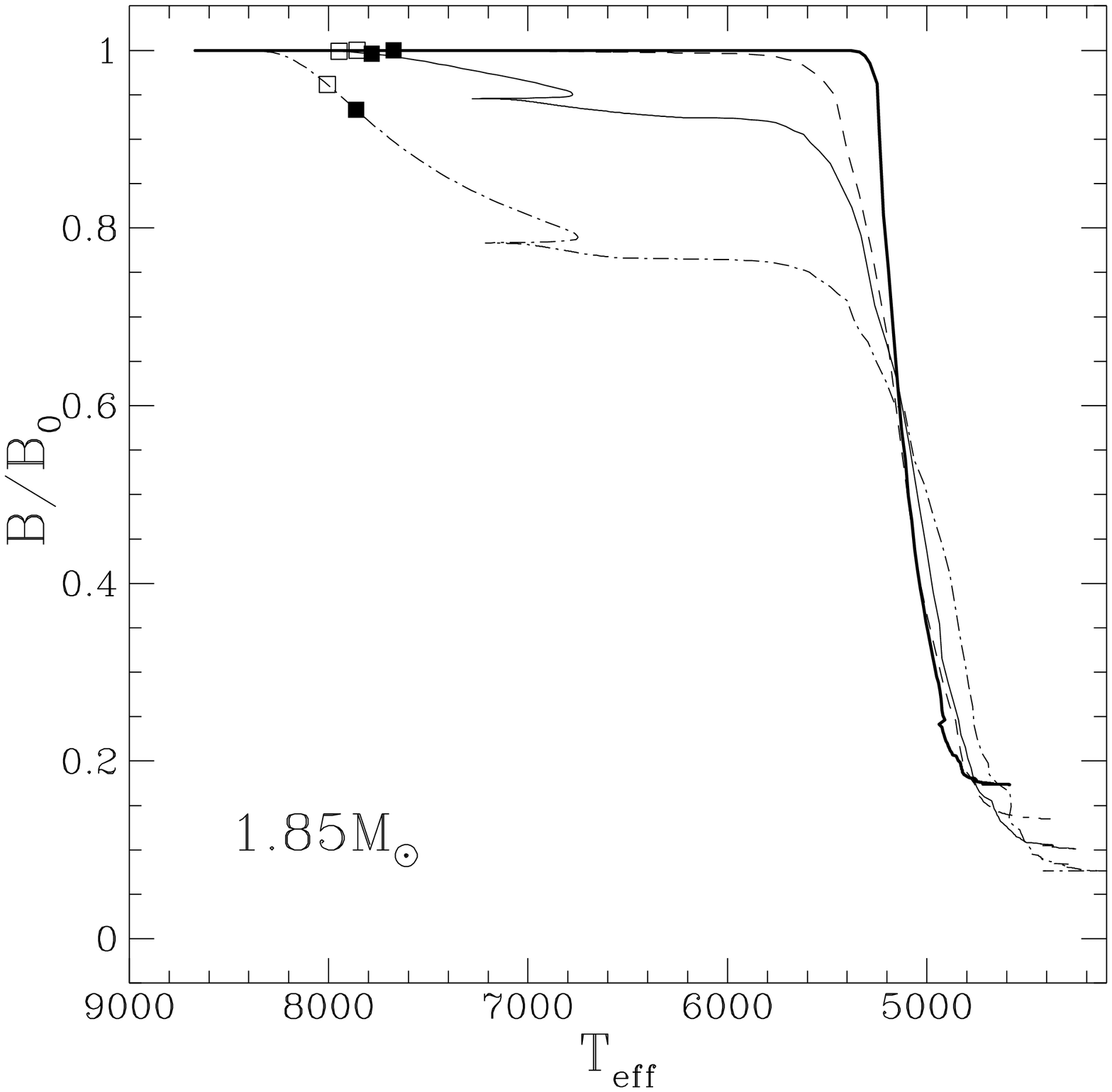,height=6cm}
}
\caption{Same as Fig.~\ref{fig_evolutionab1p5} for the 1.85 M$_{\odot}$ star
\label{fig_evolutionab1p85}}
\end{figure*}

\begin{figure*}
\centerline{
\psfig{figure=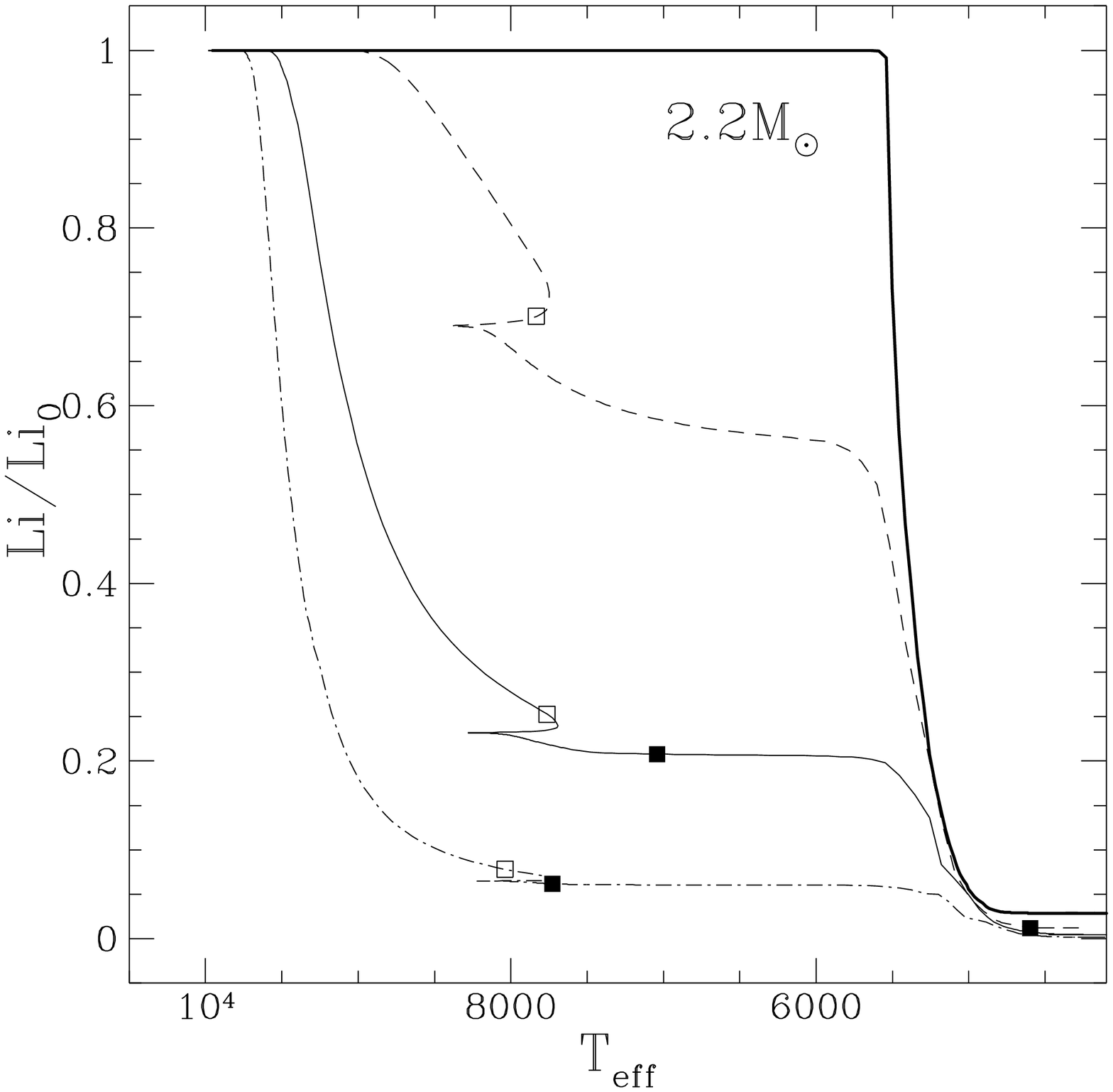,height=6cm}
\psfig{figure=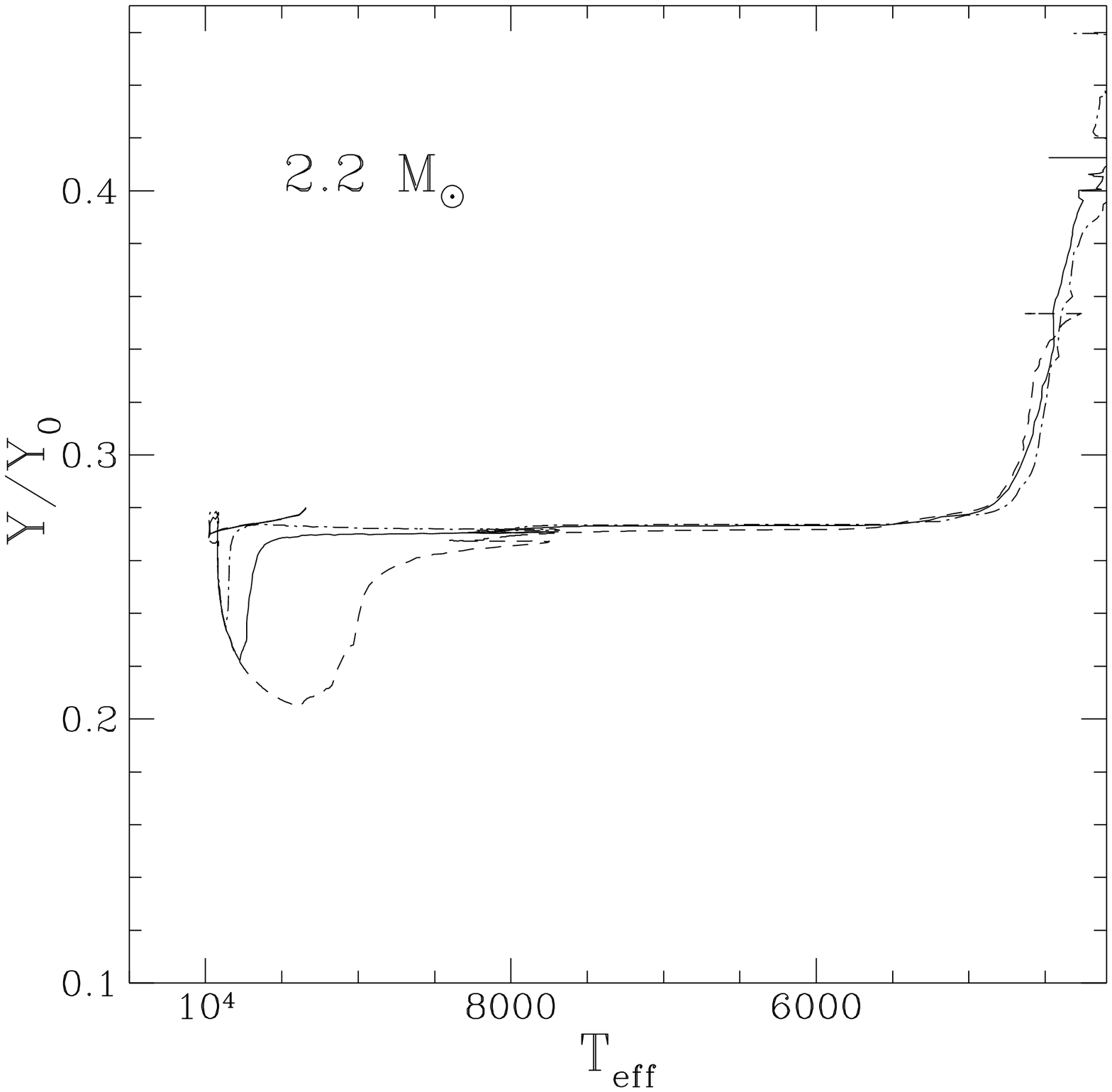,height=6cm}
}
\centerline{
\psfig{figure=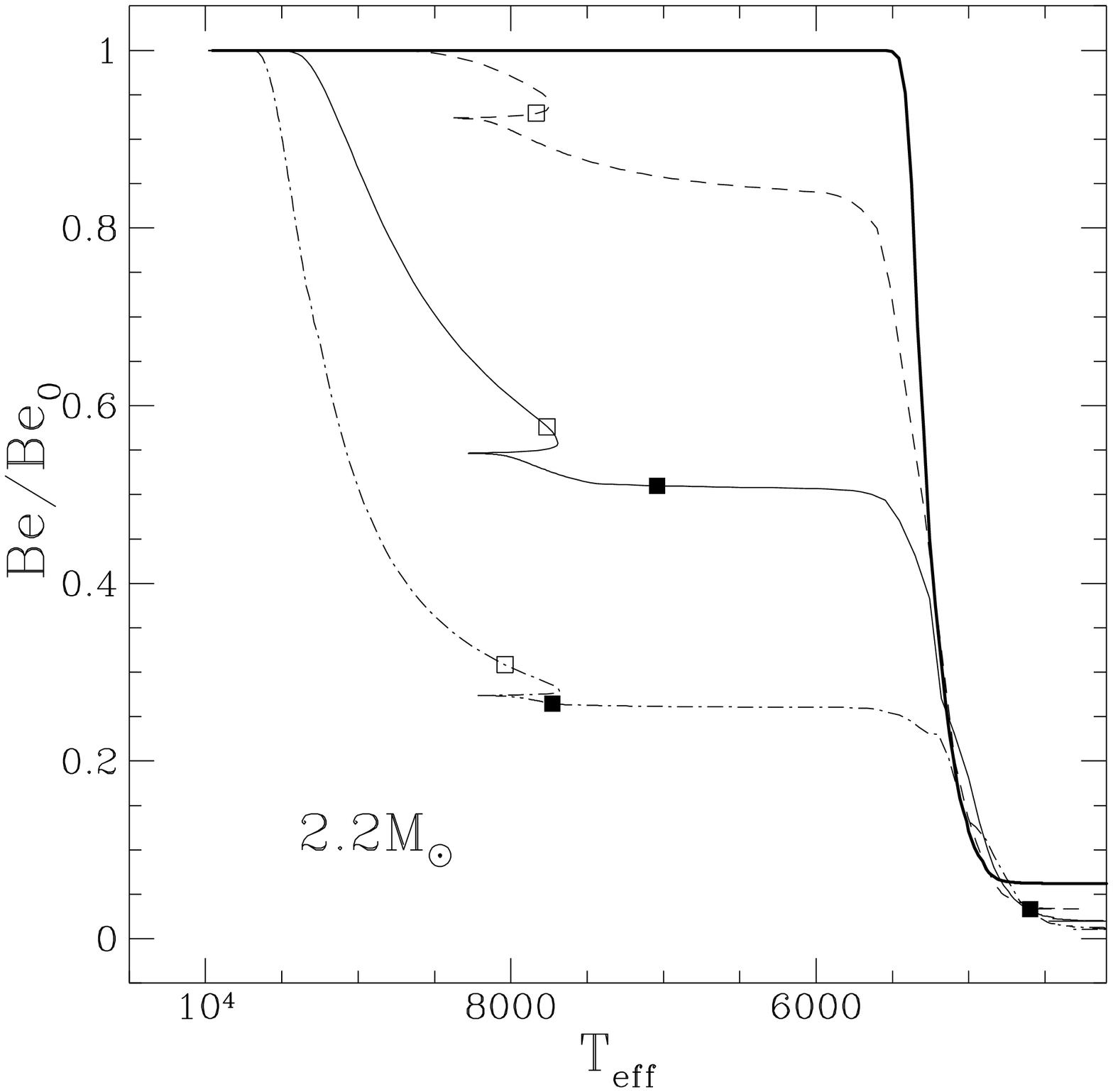,height=6cm}
\psfig{figure=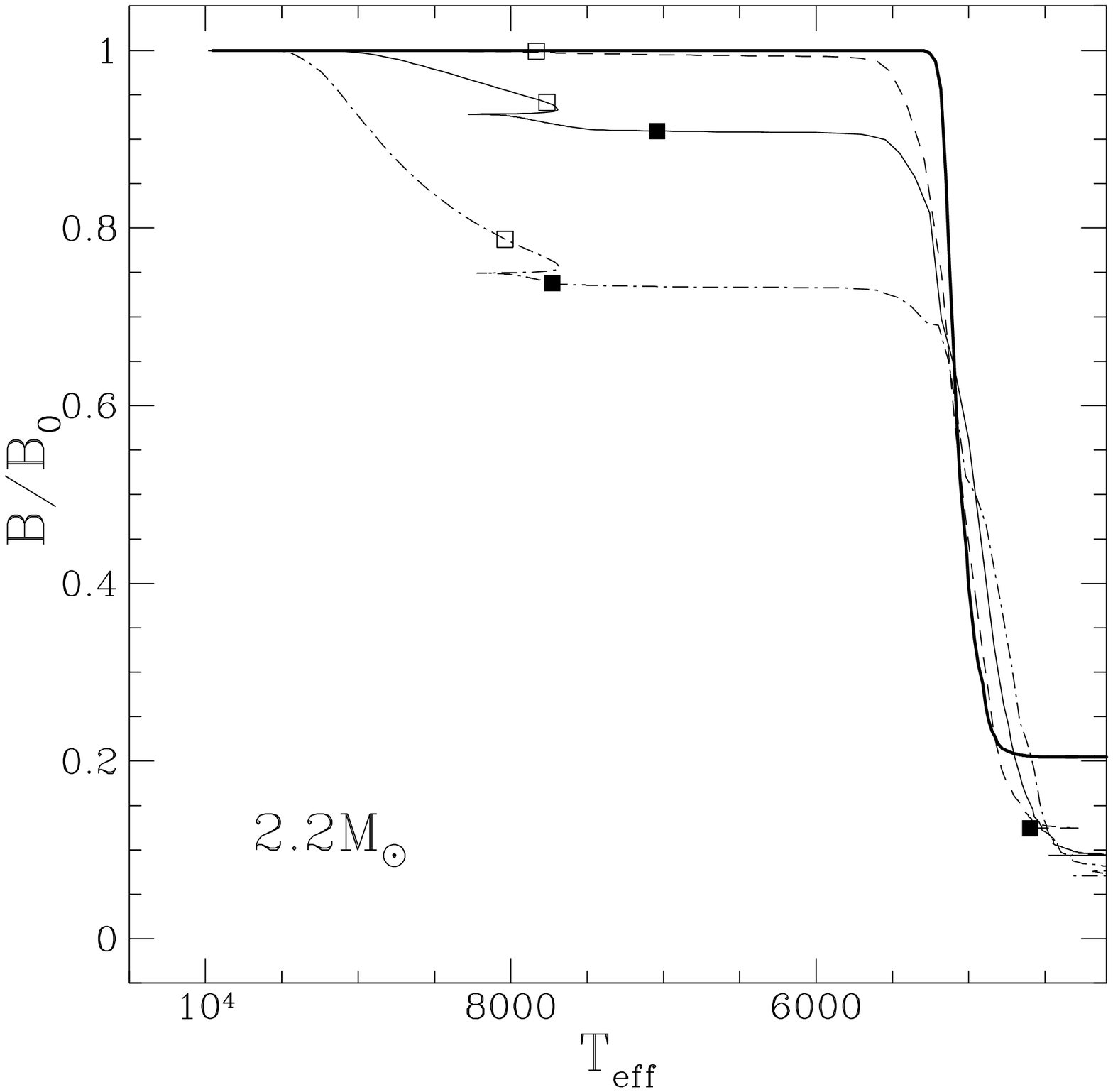,height=6cm}
}
\caption{Same as Fig.~\ref{fig_evolutionab1p5} for the 2.2 M$_{\odot}$ star
\label{fig_evolutionab2p2}}
\end{figure*}

As can be seen on the profiles of helium (Fig.~\ref{fig_abprof1p5}) 
and on the evolution of its surface abundance 
(Figs.~\ref{fig_evolutionab1p5}, \ref{fig_evolutionab1p85} and 
\ref{fig_evolutionab2p2}), the effect of microscopic diffusion is 
noticeable whatever the rotation velocity.
The mixing tends to counteract it in the faster rotators, and leads to 
smaller variations of helium in the external regions.
In the central regions, the mixing also affects the helium profile,
leading to longer main sequence lifetime when the rotation rate
is higher (see Table 1), and to slightly larger 
helium core mass.
In addition, after the turnoff the deepening convective envelope of the faster 
rotators encounters a region with more helium than in slower models and thus, 
with a lower opacity. These stars consequently evolve at a higher luminosity 
on the subgiant branch. 

\section{Evolution of the surface abundances and comparison with 
observations}

The evolution of the LiBeB and of the helium abundances at the stellar 
surface from the zero age main sequence up to the completion of the first
dredge-up can be followed in Figs.~\ref{fig_evolutionab1p5},
\ref{fig_evolutionab1p85} and \ref{fig_evolutionab2p2}. 
We indicate the location of the stars at the age of the Hyades (since the 
main sequence lifetime of the models depends on their rotation rate, 
we quote both 0.7 and 0.8 Gyr). 
We also show the surface abundance variations for models computed without 
any transport process (no rotation mixing and no atomic diffusion), in
order to locate the beginning of the standard first dredge-up.
In the models with rotation, the change of slope around 5700-5800 K is a 
signature of this event. 

\subsection{Main sequence stars}
Before discussing the LiBeB elements, one has to comment the evolution
of the helium surface abundance. In all the rotating models, the 
effect of atomic diffusion is never completely cancelled. In the very early
stages of main sequence evolution, helium settles
(cf. Figs.~\ref{fig_evolutionab1p5}, \ref{fig_evolutionab1p85} and 
\ref{fig_evolutionab2p2}). Latter on, the increase
of the macroscopic diffusion coefficients (cf. Fig.~\ref{fig_profilscd_1p5_V50V100})
leads to a partial dredge-up of helium. 
In the fast rotators (V=100, 150 km.sec$^{-1}$), this occurs 
very early on the main sequence, and the disappearance of the He$_{\rm II}$
convection zone which is necessary to explain the Am phenomenon
is not achieved. One may then find relatively few chemically 
peculiar stars with such a high rotation velocity, and the candidates would
be very young objects. 
In the slow rotators (V=50 km.sec$^{-1}$) however, the helium dredge-up occurs 
later, and the stars can appear as chemically peculiar for most of their 
main sequence life.
These predictions agree nicely with the observed velocity distribution 
of Am and normal A stars (Abt \& Morrell 1995). 

At the age of the Hyades, lithium, beryllium and boron are never depleted 
at the surface of the rotating main sequence stars on the hot side of the 
dip by more than about 
70, 40 and 5$\%$ respectively. 
The larger depletion factors in the 1.85 M$_{\odot}$ model compared to the
1.5 M$_{\odot}$ one are due to the slightly larger diffusion coefficient 
and to the smaller convection envelope that empties more rapidly. 
More important LiBeB destruction is obtained in the 2.2 M$_{\odot}$ 
model which is already evolved at the age of the Hyades. 

\begin{figure}
\centerline{
\psfig{figure=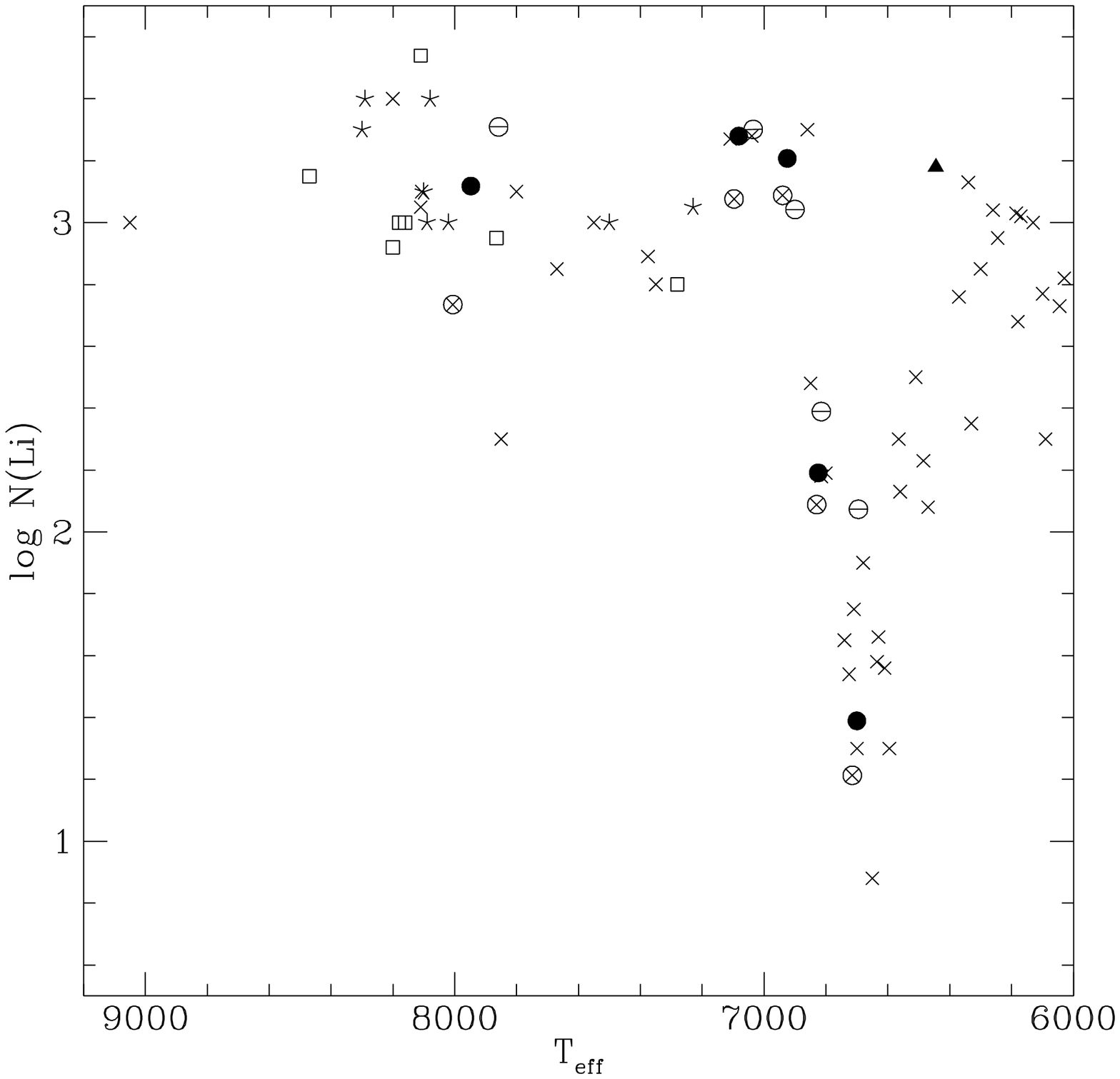,height=9cm}
}
\caption{
Lithium abundance in the Hyades (crosses), Coma (squares) and 
Praesepe (asterisks; the lithium abundances for the Praesepe stars 
with T${\rm eff}\simeq$8280 K are upper values). 
The observations are from Boesgaard \& Tripicco (1986), Thorburn et al.
(1993), Burkhart \& Coupry (1998, 1999). 
Our predictions (present paper and TC98) are also shown for comparison 
(we use logN(Li)$_0$=3.31) : 
The black dots correspond to calculations performed
with an initial velocity of 100 km.sec$^{-1}$, the pluses and minuses
surrounded by a circle correspond to models with an initial velocity of
150 and 50 km.sec$^{-1}$ respectively.
The models with T$_{\rm eff} \geq$ 6900 K conserve global angular
momentum during there main sequence lifetimes, while for the cooler
models angular momentum is lost from the surface 
\label{fig_liAmaspredictions}
}
\end{figure}

Our predictions (obtained both in this paper and in TC98)  
are compared to the observations in the Hyades, Coma and Praesepe 
in Fig.~\ref{fig_liAmaspredictions}. 
Let us recall that while the stars with T$_{\rm eff} \leq$ 6900 K lose
angular momentum from their surface, the hotter ones conserve their
global angular momentum during their main sequence life. 
In the domain we are interested in here, 
the weak differential rotation leads to diffusion 
coefficients that prevent any important surface lithium depletion 
at the age of the Hyades, in agreement with the observations. 

The abundance of the LiBeB elements keeps decreasing while stars evolve. 
At the end of the main sequence, the maximum expected lithium depletion 
in the present models is of about one order of magnitude, 
in very good agreement with the observations in field main sequence
stars originating from the hot side of the dip (Balachandran 1990,
Burkhart \& Coupry 1991).

\subsection{Evolved stars}
Even if no important lithium depletion is expected at the surface of the
main sequence stars on the hot side of the dip at the age of the Hyades, 
more lithium destruction occurs inside the rotating models compared to the 
standard case (see \S 3.3). 
This prepares the star to the lithium abundance variations observed in the 
latter evolutionary phases.
After the dredge-up, the LiBeB depletion is thus more important in the 
rotating models than in the models without transport processes (see
Table 2), due to the enlargement of the LiBeB free regions.

\begin{table}
\caption{Logarithm of the dilution factors. The observed values for Li in
different open clusters with the quoted turnoff mass are from Gilroy
(1989), while the values for field stars are from L\`ebre et al. (1999).
Be and B observed values are from Duncan et al. (1998)}
\begin{center}
\begin{tabular}{ccccc}\hline
\multicolumn{1}{c}{M$_*$/M$_{\odot}$} &
\multicolumn{1}{c}{V} &
\multicolumn{1}{c}{$^7$Li} &
\multicolumn{1}{c}{$^9$Be} &
\multicolumn{1}{c}{$^{10}$B} \\
\multicolumn{1}{c}{} &
\multicolumn{1}{c}{(km.sec$^{-1}$)} &
\multicolumn{3}{c}{Predicted} \\ \hline
1.5 & 150 & 2.87 & 1.99 & 1.11 \\
    & 100 & 2.33 & 1.68 & 0.98 \\
    &  50 & 1.89 & 1.41 & 0.84 \\
    &   0 & 1.24 & 0.93 & 0.46 \\ \hline
1.85 & 150 & 2.75 & 1.94 & 1.11 \\
     & 100 & 2.27 & 1.68 & 0.99 \\
     &  50 & 1.88 & 1.45 & 0.87 \\ \hline
     &   0 & 1.65 & 1.28 & 0.76 \\ \hline
2.2 & 150 & 2.85 & 1.97 & 1.15 \\
    & 100 & 2.34 & 1.70 & 1.04 \\
    &  50 & 1.92 & 1.48 & 0.90 \\
    &   0 & 1.54 & 1.21 & 0.70 \\ \hline
\multicolumn{1}{c}{M(turnoff)/M$_{\odot}$} &
\multicolumn{1}{c}{Open cluster} &
\multicolumn{1}{c}{$^7$Li} &
\multicolumn{1}{c}{$^9$Be} &
\multicolumn{1}{c}{$^{10}$B} \\
\multicolumn{1}{c}{} &
\multicolumn{1}{c}{} &
\multicolumn{3}{c}{Observed} \\ \hline
1.6 & NGC 752 & 1.9 - 3.2 &          &  \\
2.2 & Hyades  & 2 - 2.4 & $>$ 1.44 & 0.7 - 1.3 \\ 
2.2 & Praesepe & 2.3 - 2.7 & & \\
2.2 & IC 4756 & 2.2 - 3.3 & & \\
2.5 & NGC 6633 & 2.3 - 2.9 & & \\
2.6 & NGC 2548 & 2.3 - 2.6 & & \\
2.7 & NGC 2281 & 3.1 - 3.2 & & \\
2.8 & Stock 2  & 2.5 - 3 & & \\
2.8 & NGC 1545 & $>$ 3 & & \\
1.5 - 3 & Field & 2.1 - $>$ 3 & & \\ \hline
\end{tabular}
\end{center}
\end{table}

The predicted evolution of the surface lithium abundance in our rotating
models explains the 
behavior in the stars more massive than
1.5 M$_{\odot}$ observed by L\`ebre et al. (1999) and discussed by Dias
et al. (1999). Indeed, for these objects a lithium dispersion of about
two orders of magnitude is seen already at T$_{\rm eff} \simeq$ 6300 K, 
i.e. before the beginning of the standard dilution (at T$_{\rm eff}
\simeq$ 5600 K). In the same
homogeneous sample, the observed dilution factors are higher than the ones 
predicted by standard dilution alone. 
This is also seen in open cluster giants with turnoff masses higher than
1.5 M$_{\odot}$ (Gilroy 1989)
and in the large sample of field giants of Brown et al. (1989).
As can be seen in Table 2, lithium depletion in giants down to factors
as large as 1000 is well reproduced within our framework. 

The effect is especially important for Li, which is
the more fragile element. Regarding Be and B, the post-dilution values 
decrease at most by a factor of 3 (4) and 5 (11) respectively in 
the 2.2 (1.5) M$_{\odot}$ model compared to the case without mixing. 
From HST spectra, Duncan et al. (1998) determined a dilution factor 
of 1.0$\pm$0.3 dex for B in two Hyades giants which present a lithium
dilution factor of 2.2$\pm$0.2 (see Table 2). 
These data are nicely explained by our rotating models. 

\section{Conclusions}
In TC98, we presented stellar models including the most complete
description currently available for rotation induced-mixing, taking into
account simultaneously the transport of chemicals and the transport of
angular momentum due to the wind-driven meridional circulation. 
We showed that the shape of the hot side of the Li dip in open clusters 
is well explained within this framework, which also successfully
reproduces the C and N anomalies in B type stars (Talon et al. 1997). 

In the present paper, we studied the impact of rotational mixing in A
and early-F type stars on the main sequence and up to the completion of 
the first dredge-up. 
We showed that low lithium abundances measured in evolved stars originating 
from the hot side of the dip, which can not be explained by standard
dilution alone, can be linked simply to destruction in the interior of 
stars still on the main sequence.
Such a destruction is not visible at the surface of the Hyades main sequence 
stars with T$_{\rm eff} \geq$ 6900 K simply because they are too young.
More BeB data are needed to check the validity of our predictions for
these elements.

\begin{acknowledgements}
We would like to thank Claude Burkhart for sending us data prior to
publication.
S.T. gratefully acknowledges support from FCAR of Quebec and
NSERC of Canada.
\end{acknowledgements}

\appendix

\end{document}